\documentclass[openacc]{rstransa}

\usepackage{graphics}
\usepackage{graphicx}

\usepackage[numbers,sort&compress]{natbib}

\newcommand*{\doi}[1]{\href{http://dx.doi.org/#1}{doi: #1}}

\usepackage{tcolorbox}
\definecolor{abstract-color}{cmyk}{0.04, 0.04, 0.12, 0.08}

\begin{document}

\title{Spectropolarimetric Fluctuations in a Sunspot Chromosphere}

\author{M. Stangalini$^{1,2}$, D. Baker$^{3}$, G. Valori$^{3}$, \\ D.B. Jess$^{4,5}$, S. Jafarzadeh$^{6,7}$, \\ M. Murabito$^{2}$, A.S.H. To$^{3}$, D.H. Brooks$^{8}$, \\ I. Ermolli$^{2}$, F. Giorgi$^{2}$, C.D. MacBride$^{4}$}

\address{$^{1}$ASI, Italian Space Agency, Via del Politecnico snc, 00133 Rome, Italy\\
$^{2}$ INAF-OAR National Institute for Astrophysics, 00078 Monte Porzio Catone (RM), Italy\\
$^{3}$ University College London, Mullard Space Science Laboratory, Holmbury St. Mary, Dorking, Surrey, RH5 6NT, UK  \\
$^{4}$Astrophysics Research Centre, School of Mathematics and Physics, Queen's University Belfast, Belfast, BT7 1NN, U.K. \\
$^{5}$Department of Physics and Astronomy, California State University Northridge, Northridge, CA 91330, U.S.A. \\
$^{6}$Rosseland Centre for Solar Physics, University of Oslo, P.O. Box 1029 Blindern, NO-0315 Oslo, Norway\\
$^{7}$Institute of Theoretical Astrophysics, University of Oslo, P.O. Box 1029 Blindern, NO-0315 Oslo, Norway\\
$^{8}$ College of Science, George Mason University, 4400 University Drive, Fairfax, VA 22030, USA}
\subject{astrophysics, solar physics, MHD waves}


\corres{Marco Stangalini\\
\email{marco.stangalini@asi.it}}

\maketitle

\begin{tcolorbox}[sharp corners, width=\textwidth,colback=abstract-color,colframe=abstract-color,boxsep=5pt,left=0pt,right=0pt,top=0pt,bottom=0pt]
The instrumental advances made in this new era of 4-meter class solar telescopes with unmatched spectropolarimetric accuracy and sensitivity, will enable the study of chromospheric magnetic fields and their dynamics with unprecedented detail. 
In this regard, spectropolarimetric diagnostics can provide invaluable insight into magneto-hydrodynamic (MHD) wave processes. 
MHD waves and, in particular, Alfv\'enic fluctuations associated to particular wave modes, were recently recognized as important mechanisms not only for the heating of the outer layers of the Sun's atmosphere and the acceleration of the solar wind, but also for the elemental abundance anomaly observed in the corona of the Sun and other Sun-like stars (also known as first ionisation potential; FIP) effect.
Here, we take advantage of state-of-the-art and unique spectropolarimetric IBIS observations to investigate the relation between intensity and circular polarisation ($CP$) fluctuations in a sunspot chromosphere. Our results show a clear link between the intensity and $CP$ fluctuations in a patch which corresponds to a narrow range of magnetic field inclinations. This suggests the presence of Alfv\'enic perturbations in the sunspot.
\end{tcolorbox}



\section{Introduction}
A large variety of different magnetohydrodynamic (MHD) wave modes are theoretically predicted, and indeed observed, in solar magnetic structures ranging from large sunspots extending over several Mm, down to very small magnetic elements at the limit of spatial resolution on current solar telescopes \cite{Edwin1983, Roberts1983, Khomenko2008, 2010ApJ...719..357F, 2011ApJ...727...17F, 2015MNRAS.449.1679M, 2015ApJ...799....6M, 2016PhDT........15L, Grant:2018vfa, 2018ApJ...857...28K}.
Magnetic fields connect different heights in the solar atmosphere, thus MHD waves propagating along them \cite{2006ApJ...640.1153C} can play a major role in supplying the energy budget to the upper layers of the solar atmosphere \cite{2009Jess, 2011ApJ...735...65F, 2012NatCo...3E1315M, 2013SSRv..175....1M, 2015SSRv..190..103J, 2017ApJ...847....5K}. 
When particular magnetic field geometries and local physical conditions are met, propagating MHD waves of one type can be converted into another (e.g. from slow to fast magneto-acoustic modes and/or Alfv{\'{e}}n waves \cite{2006ApJ...647L..77M, 2006ApJ...648L.151J, 2007ApJ...671.1005B, 2011A&A...534A..65S, 2015A&A...579A..73M, 2017SoPh..292...35A, 2018ApJ...853..136K}).
Observationally, MHD waves in the solar atmosphere are commonly identified as intensity and velocity oscillations \cite{2000SoPh..192..373B, 2006ApJ...640.1153C, 2010A&A...513A..27C, 2011ApJ...729L..18M, 2012A&A...539L...4S, 2015ApJ...806..132G, 2017ApJS..229...10J, 2017ApJ...842...59J}, although associated magnetic field oscillations are also expected from theory \cite{Edwin1983, Roberts1983}. These quantities may have different phase relations depending on the MHD mode and the propagation state of the wave \cite{2009ApJ...702.1443F, Moreels2013, MoreelsGoossens, 10.1093/pasj/psz084}.
\begin{figure}[!t]
    \centering
    \includegraphics[trim=0 0 4 0,clip,width=\textwidth]{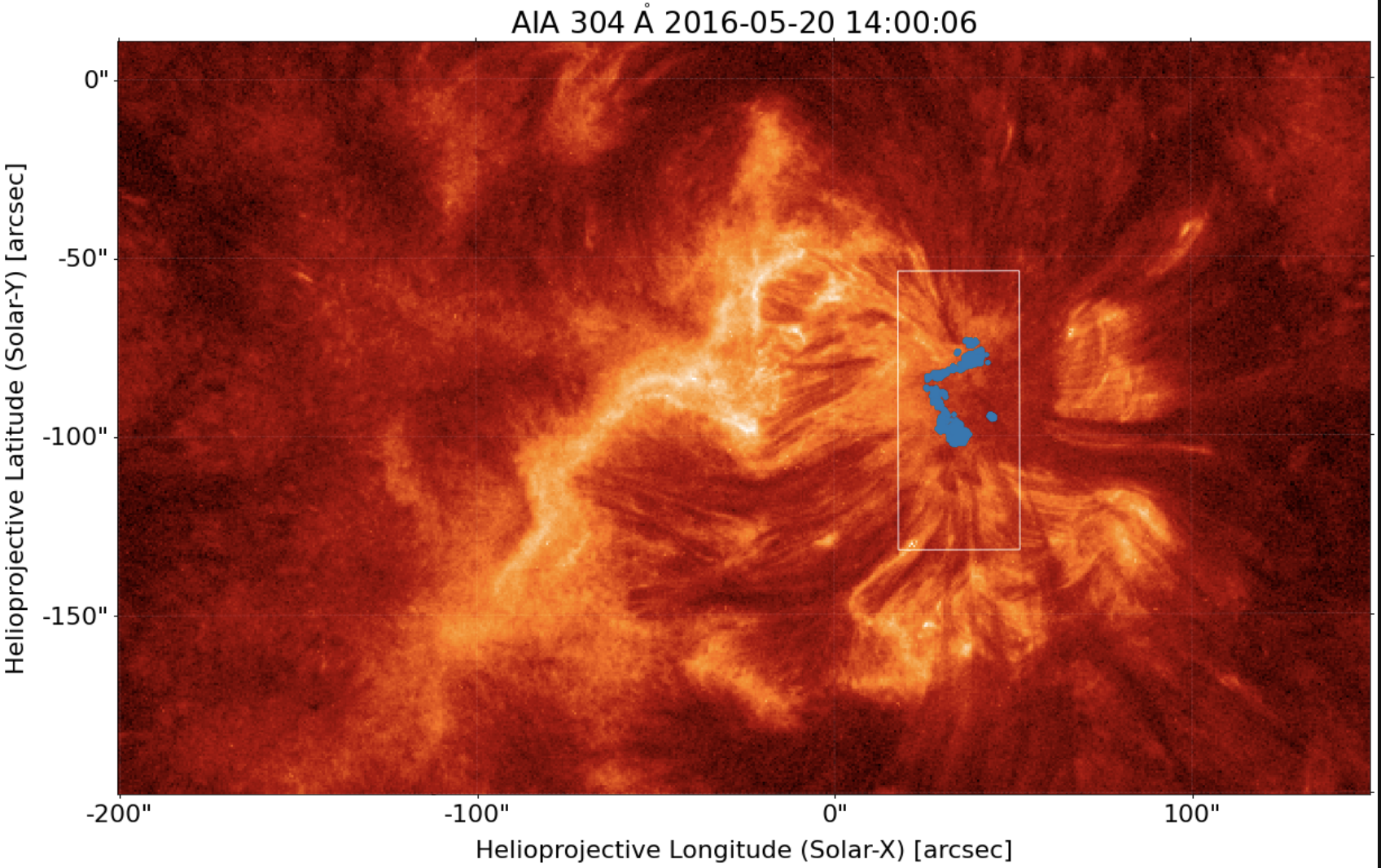}   
    \includegraphics[trim=90 0 80 0,clip,width=\textwidth]{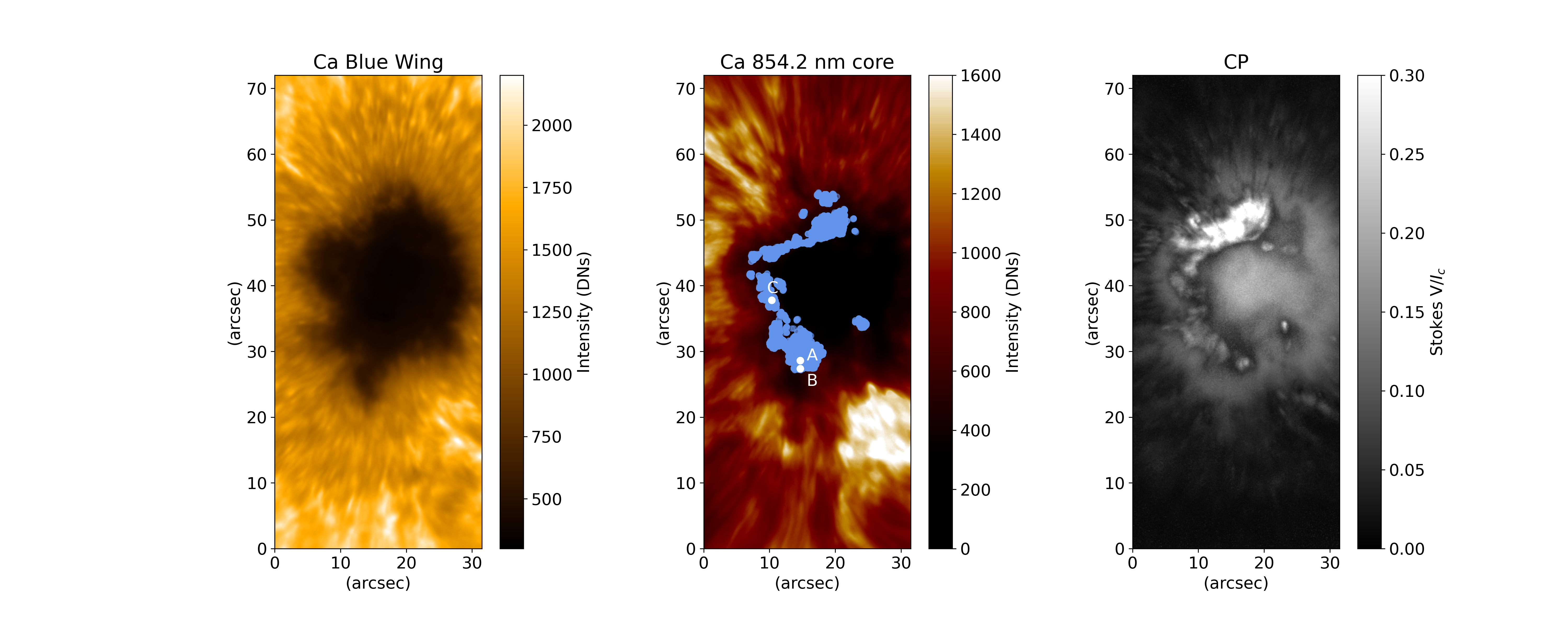}
    \caption{Upper panel: SDO/AIA He~{\sc{ii}}~$30.4$~nm context image of AR  12546. Bottom left: IBIS narrow band intensity image in the blue wing of the Ca~{\sc{ii}}~$854.2$~nm spectral line. Bottom center: IBIS intensity image in the core of the Ca~{\sc{ii}}~$854.2$~nm spectral line. The blue dots indicate the positions where there exists a large ($>95\%$) coherence between the intensity and $CP$ fluctuations (see text). The three white dots indicate the positions corresponding to the wavelet diagrams shown in Fig.~{\ref{fig:2}}.  Bottom right: Ca~{\sc{ii}}~854.2~nm circular polarisation map. }
    \label{fig:1}
\end{figure}
It is worth noting that the relevance of magnetic field perturbations goes well beyond the MHD wave field. Recently, it was shown that magnetic perturbations may give rise to, through the generation of the ponderomotive force, the well-known FIP (first ionization potential) effect, an elemental abundance anomaly observed in the corona of the Sun and other Sun-like stars \cite{2015LRSP...12....2L}.
In recent years, and most notably with the in-situ observations expected from the Solar Orbiter mission \cite{2013SoPh..285...25M}, the FIP effect has acquired new relevance as a powerful diagnostic of the solar wind, especially useful for linking the solar wind to its source regions \cite{2019ApJ...879..124L, 10.1093/pasj/psz084}.\\
Many authors have reported \cite{1997SoPh..172...69H, Norton_1999, 1999ASSL..243..337R, 1999SoPh..187..389B, 2002AN....323..317S, 2000ApJ...534..989B} magnetic field oscillations with periods in the range $3-5$~min and oscillation amplitudes ranging from 10~G to a few hundreds G in sunspots, concentrated either in patches \cite{1998A&A...335L..97R}, or at the umbra-penumbra boundary \cite{1999SoPh..187..389B, 2000SoPh..191...97K}.
Very recently, extremely long period (up to $60$ hours) magnetic oscillations were also reported in a number of sunspots \citep{2020A&A...635A..64G}.
However, the interpretation of these fluctuations in terms of intrinsic magnetic oscillations is debated \cite{1998ApJ...497..464L, 2015LRSP...12....6K}. Spurious magnetic oscillations may in fact arise from the cross-talk between other physical quantities. In particular, in the presence of a vertical gradient of the magnetic field (as is the case for many magnetic structures in the solar atmosphere), opacity effects can also play a significant role \cite{2015LRSP...12....6K}. Indeed, due to opacity changes, the height of formation of the spectral line changes too, hence the sampled height varies from which the information is extracted.
Since the magnetic field is not constant, such variable height sampling results in apparent fluctuations of the magnetic flux.
A telltale sign of such a situation arising from opacity effects is that the apparent magnetic field fluctuations are in phase (or out-of-phase) with density fluctuations (or diagnostics derived thereof; \cite{2009ApJ...702.1443F}).
Analyses of the phase relations between magnetic field diagnostics, such as the circular polarisation ($CP$) of the spectral line, and other independent diagnostics (e.g. intensity, Doppler velocity, etc.) acquired at two atmospheric heights are useful to distinguish real magnetic oscillations from opacity effects and cross-talk \cite{1999SoPh..187..389B, 2000SoPh..191...97K, 2009ApJ...702.1443F, 2015A&A...579A..73M}, and to correctly identify different MHD modes \cite{2003A&A...410.1023R, 2011ApJ...730L..37M, 2013A&A...551A.137M, 2013A&A...555A..75M, 2015A&A...579A..73M}.
An important aspect of phase lag analysis is that only phase measurements with large coherences should be considered in order to ensure the reliability of the results. Concerning this, seeing effects can easily destroy the coherence between different diagnostics and hamper the disambiguation of magnetic oscillations from other oscillatory effects. For this reason, long data sequences acquired from space or from the ground with adaptive optics (AO) systems under good and stable seeing conditions are required.

As of now, most of the literature on the subject is focused on photospheric data (see for instance \citet{2020A&A...635A..64G}). However, with high sensitivity spectropolarimetry progressively extending towards the upper layers of the Sun's atmosphere, it is now becoming possible to examine the chromospheric response of spectropolarimetric diagnostics to wave dynamics, enabling new possibilities in the identification of the various modes and, possibly, their dissipation or conversion. These capabilities will be even more enhanced with the advent of new four meter class solar telescopes like DKIST and EST, which will provide spectropolarimetric observations of the solar chromosphere with unmatched spatial resolution and polarimetric precision.
In this regard, \citet{2018A&A...619A..63J} have shown, through state-of-the-art spectropolarimetric observations at chromospheric heights, the presence of Alfv{\'{e}}nic disturbances not consistent with opacity effects in the chromosphere of a large sunspot. These results were obtained by employing sophisticated spectropolarimetric inversion techniques, which represent a viable alternative approach to solving any opacity issues.

To demonstrate the rich potential of spectropolarimetric diagnostics for MHD wave studies in the chromosphere, in this work we take advantage of state-of-the-art and unique observations of one of the biggest sunspots of the last twenty years acquired by IBIS, the Interferometric BIdimensional Spectrometer (IBIS; \cite{MScavallini06, MSreardon08}) instrument at the National Science Foundation's Dunn Solar Telescope, to explore the relation between spectropolarimetric diagnostics and, in particular, circular polarisation and intensity oscillations.

\section{Data and Method}
The data used in this study consist of an exceptionally long (184~min) sequence of high-spatial (0.2~arcsec) and temporal (48~s) resolution spectropolarimetric observations of AR12546, obtained with IBIS on 2016 May 20, starting at 13:39~UTC. The observations consist of spectropolarimetric spectral scans (21 spectral points per scan) of the  Ca~{\sc{ii}} 854.2~nm absorption line, acquired under excellent and stable seeing conditions. These uncommon circumstances for ground based observations, together with the exceptional size of the sunspot (see Fig.~{\ref{fig:1}}), make this data set unique and best suited for wave studies. At the start of the data sequence, AR12546 was close to disk centre at a location of [$ 7^\circ$ S, $2^\circ$ W].
The observations were assisted by the high-order AO system \cite{2004SPIE.5490...34R}, which was in stable closed-loop conditions for all the duration of the observation. The integration time was set to $80$~ms per exposure. 
In addition, the calibrated images were restored with multi-object multi-frame blind deconvolution (MOMFBD; \cite{MSnoort05}) to limit the effect of residual seeing aberrations.

This study is based on the circular polarisation signals on a pixel-by-pixel basis, which is computed as the amplitude of the Stokes-$V$ profile:
\begin{equation}
CP = \frac{|V_{\text{max}}|}{I_{\text{cont}}} \cdot \text{sign}(V_{\text{max}}) \ ,
\end{equation}
where $V_{\text{max}}$ is the maximum amplitude of the Stokes-$V$ spectral profile, and $I_{\text{cont}}$ the local continuum intensity. This is done for all the scans of the data sequence.
The spectropolarimetric sensitivity of IBIS was estimated in \cite{2010ApJ...723..787V} as $10^{-3}$ of the continuum intensity level.  In addition to radiometric calibration, polarimetric demodulation and cross-talk minimization between the Stokes parameters was included in the calibration process. This data set was also studied in \citet{Stangalini_2018, 2019ApJ...873..126M, 2020ApJ...892...49H, 2020ApJ...890...96M} 

\begin{figure}[!t]
    \centering
    \includegraphics[trim=10 0 0 0,clip,scale=0.45]{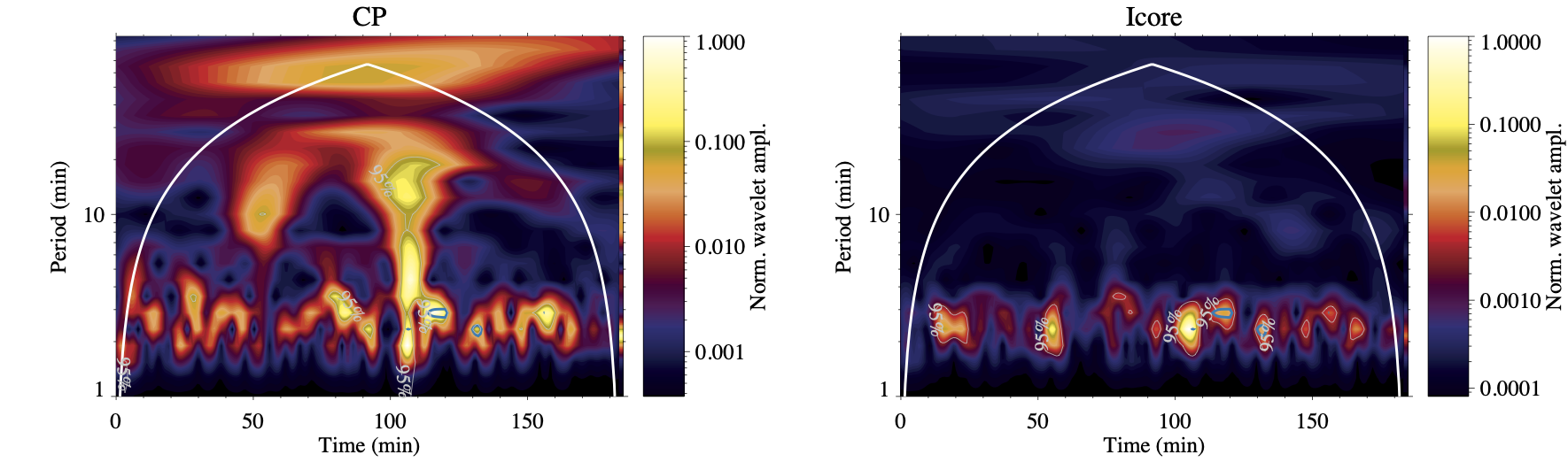}
    \includegraphics[trim=10 0 0 0,clip,scale=0.45]{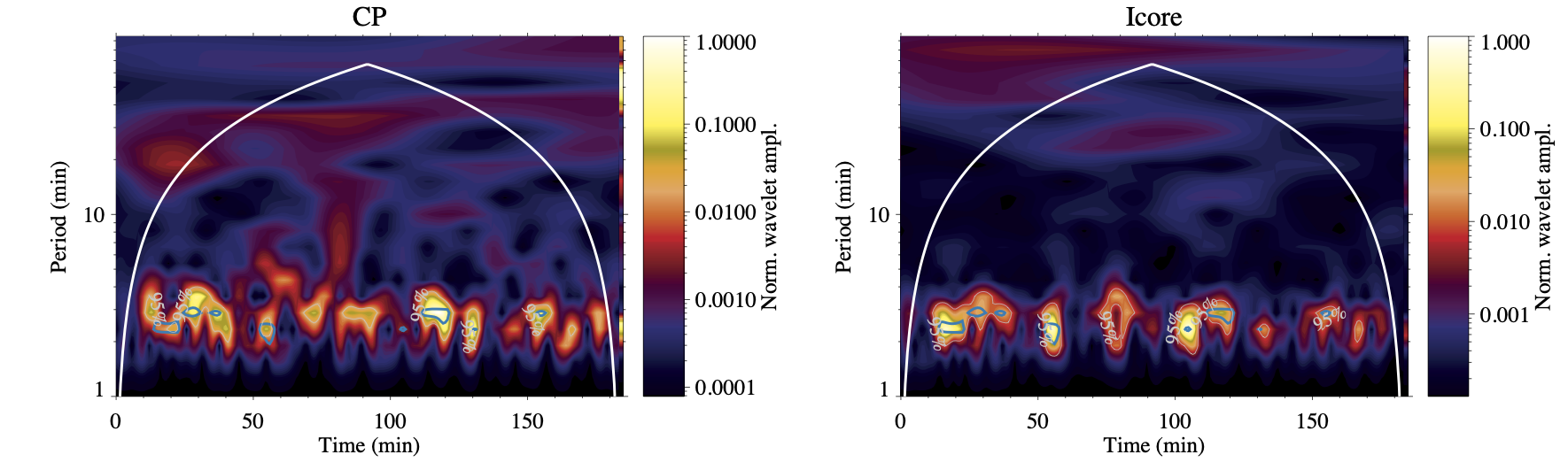}
    \includegraphics[trim=10 0 0 0,clip,scale=0.45]{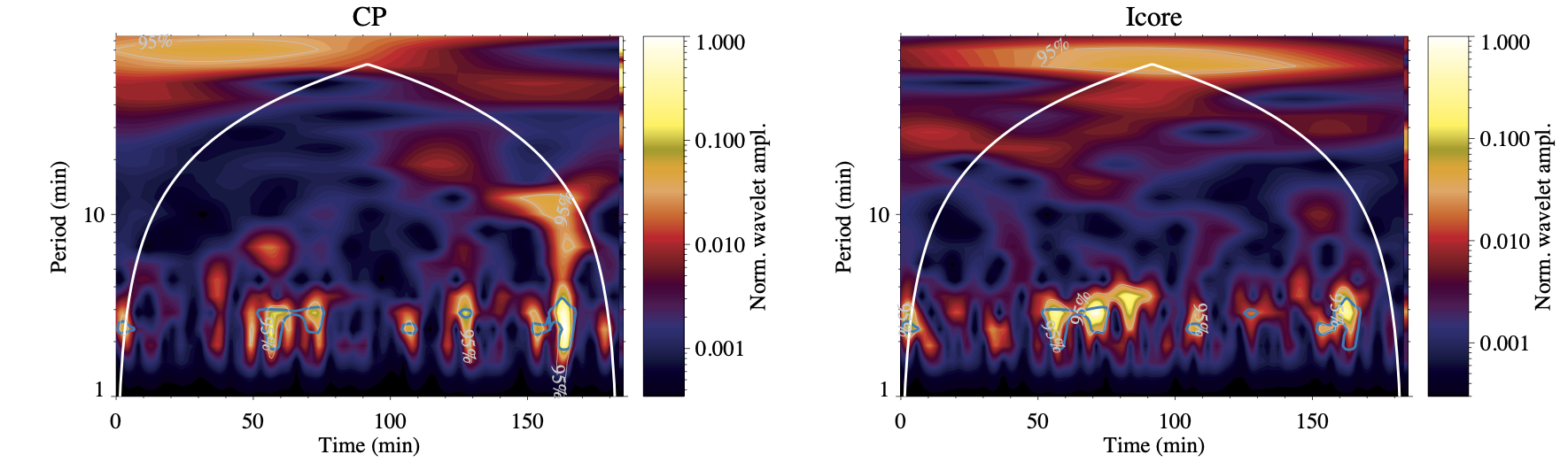}    
    \caption{Time-period wavelet power spectrum of the $CP$ (left column) and intensity (right column) signals at three positions A, B, and C shown in Fig. \ref{fig:1} (from top to bottom).The white contour represents the 95\% statistical confidence level of the wavelet analysis, while the blue contours the coherence between CP and intensity oscillations larger than $90\%$.}
    \label{fig:2}
\end{figure}

\begin{figure}[!t]
    \centering
    \includegraphics[trim=100 0 0 10,clip,width=12cm]{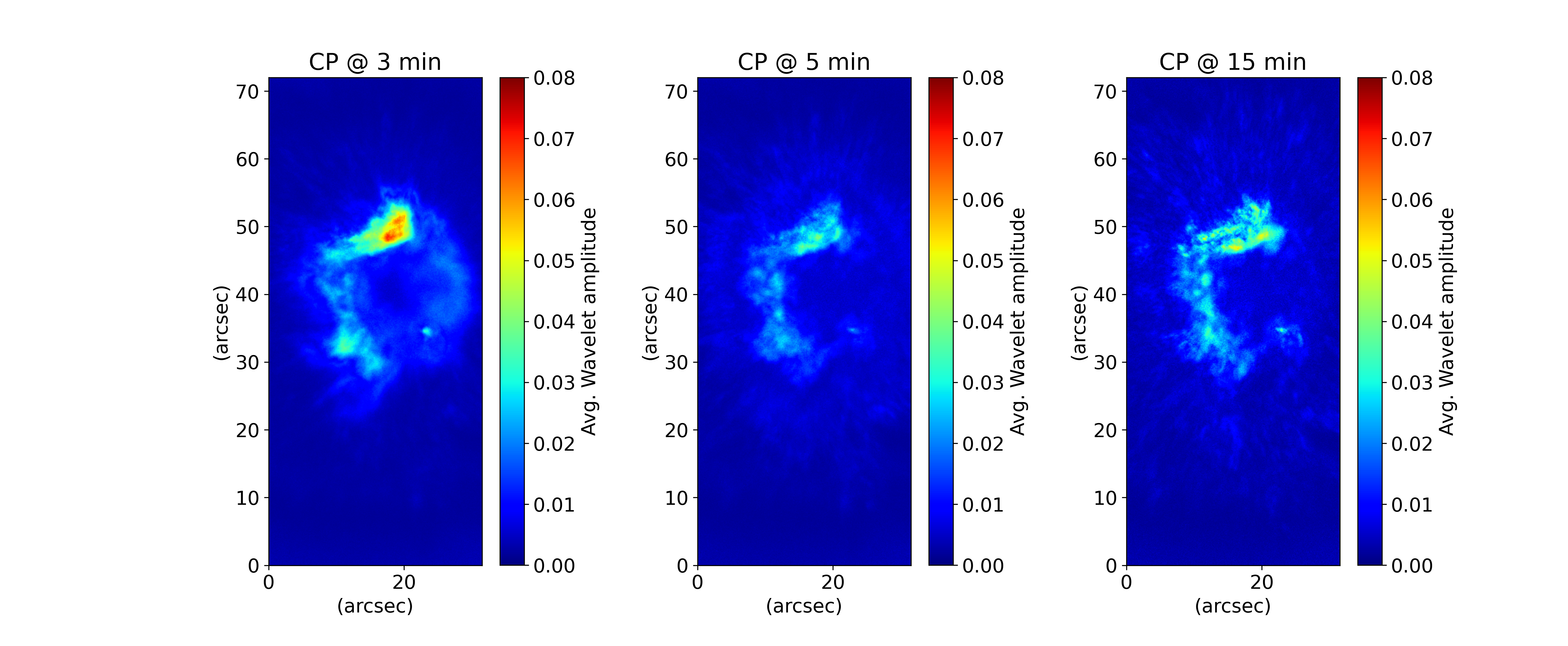}
    \includegraphics[trim=100 0 0 10,clip,width=12cm]{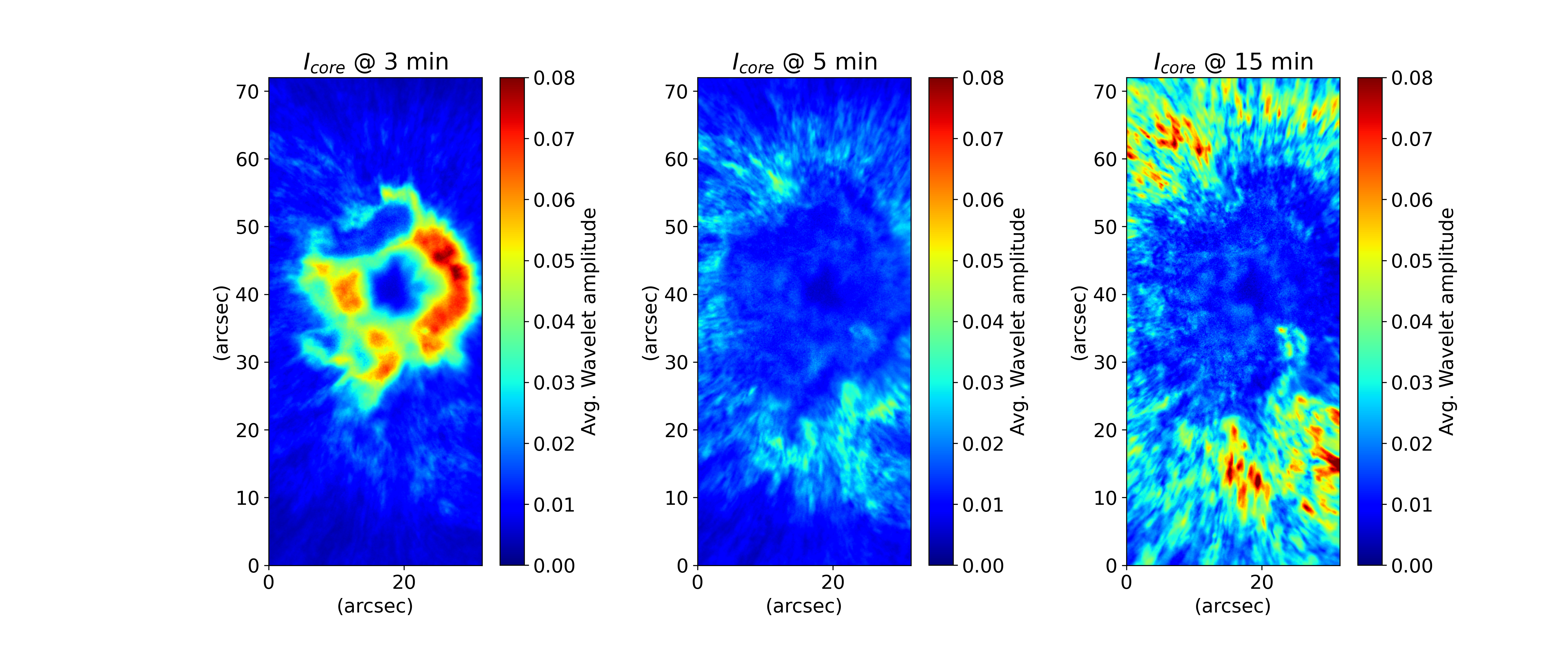}
    \caption{Temporally averaged wavelet amplitude in the 3, 5, and 15 minute range, width 1 min, for $CP$ (top) and core intensity (bottom).}
    \label{fig:3}
\end{figure}

In order to ensure the reliability of the results, here we restrict our attention to the intensity and $CP$ fluctuations only. 
This choice is also motivated by the fact that since the Ca~{\sc{ii}}~$854.2$~nm spectral line often goes into emission, the estimation of the Doppler velocity may lead to inconsistencies \citep{2020arXiv200707904M}. Inversion techniques may provide a solid context for the reliable estimation of the Doppler velocity even in these conditions, but such a technique will be addressed and explored in future work on the subject.
The focal point of our investigation is the wavelet analysis of the $CP$ and intensity signals at each pixel in our field of view (FoV). The signals associated with MHD waves can be highly non-stationary \cite{Stangalini2012}, thus wavelet analysis is a more suitable methodology to employ in this case.
To this aim, we made use of the well-known wavelet analysis code described in \citet{Torrence1999}, and already employed in many other works focusing on waves in the solar atmosphere (see for instance \citet{2013AA...554A.115S, 2017ApJS..229...10J}), and making use of the standard Morlet mother function to investigate the appearance of power features in the period-time wavelet diagram.\\ 
Finally, we focus on the phase relations between the two signals that, as discussed before, can be important information to disentangle real magnetic oscillations (if any) from opacity effects and cross-talk. To this aim, we  extract the phase information between the $CP$ and intensity signals from the wavelet coherence, which is obtained from the same code. 
In order to ensure the reliability of the results,  we only considered phase measurements corresponding to a coherence larger than $90\%$.
It is worth emphasising that a large coherence is not necessarily associated with a large amplitude oscillation in the two signals at the given frequency/period (i.e. large power). In other words the coherence quantifies the degree of dependence of the two signals independently of their amplitude.

\section{Results}
\subsection{Wavelet Analysis}
In Fig.~\ref{fig:2} we show, as an example, the wavelet diagrams of the $CP$ and the intensity fluctuations at the three positions (A, B, and C) indicated in the bottom center panel of Fig.~\ref{fig:1}. The $95\%$ confidence level contours (white lines) highlight the statistical significance of these oscillations. 
The three positions are located where a particular phase relation between CP and intensity suggests the presence of intrinsic magnetic oscillations. In addition, these positions have wavelet diagrams which are representative of those obtained in correspondence of all the blue dots in Figure \ref{fig:1}, and are therefore chosen to highlight particular aspects of the oscillations seen at those locations.
The first thing to note here is that the power of the fluctuations in both intensity and $CP$ is mainly concentrated in a band centered around 3~minutes. Moreover, the oscillations at different locations are not in phase (see for example oscillations at position A and C).  This suggests that local oscillations play a major role. Another important aspect is that, while in the 3-minute band there is a quite good correspondence between CP and intensity, CP wavelet diagrams occasionally also present power features at longer periods, up to $10-15$ min (see top and bottom panels). These features have significance larger than $95\%$, and are found outside the cone of influence  represented by the continuous white line, thus they are reliable oscillations; inside the cone of influence   oscillations cannot be trusted due to the limited length of the time series. Since we are interested in identifying intrinsic magnetic oscillations, long period CP oscillations appear particularly relevant here since no intensity oscillations are found, as we would expect for opacity effects. It is worth recalling here that the long duration of our data sequence (more than 3 hours) is well suited to the search of long period oscillations as highlighted by the cone of influence in the plots.\\ 
In order to better investigate the relation between intensity and CP fluctuations, it is useful to study the spatial distribution of their power.
In Fig.~\ref{fig:3} we show the average wavelet amplitude maps in the $3$, $5$, and $15$ min bands (each one averaging over a $1$~min width).
The value at each pixel in this figure is constructed by taking the time-average (horizontal direction in Fig.~\ref{fig:2}) of the period-time wavelet diagram in each of the three period bands.
Here we note that $CP$ and intensity do not share a similar spatial distribution of their oscillatory power.
The intensity at all bands show, in fact, a peculiar circular pattern of fluctuations with long period oscillations (i.e. 5 and 15 min period) dominating the penumbra and 3-minute oscillations concentrated in the umbra. On the other hand, $CP$ oscillations are mainly localised in a \textit{C-shaped} region, and this is the case for all bands.
However, from the wavelet diagrams in Fig.~\ref{fig:2}, the highest power of the oscillations is mainly concentrated in the 3~minute band.\\
Regarding the lack of intensity oscillations in the central and upper-left regions of the umbra, we note that this effect could be due to one or a combination of factors. Firstly, opacity effects, by lowering the height of formation of the spectral line, allow the sampling of intrinsically lower layers of the Sun's atmosphere which, due to the atmospheric stratification, display smaller oscillation amplitudes. Therefore, in this case, one should expect an increase of the $CP$ signal.  However, this is observed only in the upper-left part of the umbra, not at it's center, thus opacity effects alone cannot justify the lack of intensity power. It is worth recalling that the extremely intense magnetic field at the center of the umbra ($\sim$ 4 kG) can significantly reduce the acoustic power at the same location (see for instance \citep{2012AA...539L...4S}), resulting in a reduction of power also at chromospheric heights. Finally, \citet{Stangalini_2018}  also found evidence of a surface mode in the same sunspot between the photosphere and the chromosphere. This may may also contribute to the increase of power towards the edge of the umbra. while all these effects or a combination of them may lead to the observed pattern, it is certainly difficult to judge which is the main contributor. Spectropolarimetric inversions may provide essential information in this regard in the future.

\subsection{Phase Lag Analysis}
\begin{figure}[!t]
    \centering
    \includegraphics[trim=10 0 0 0,clip,scale=0.3]{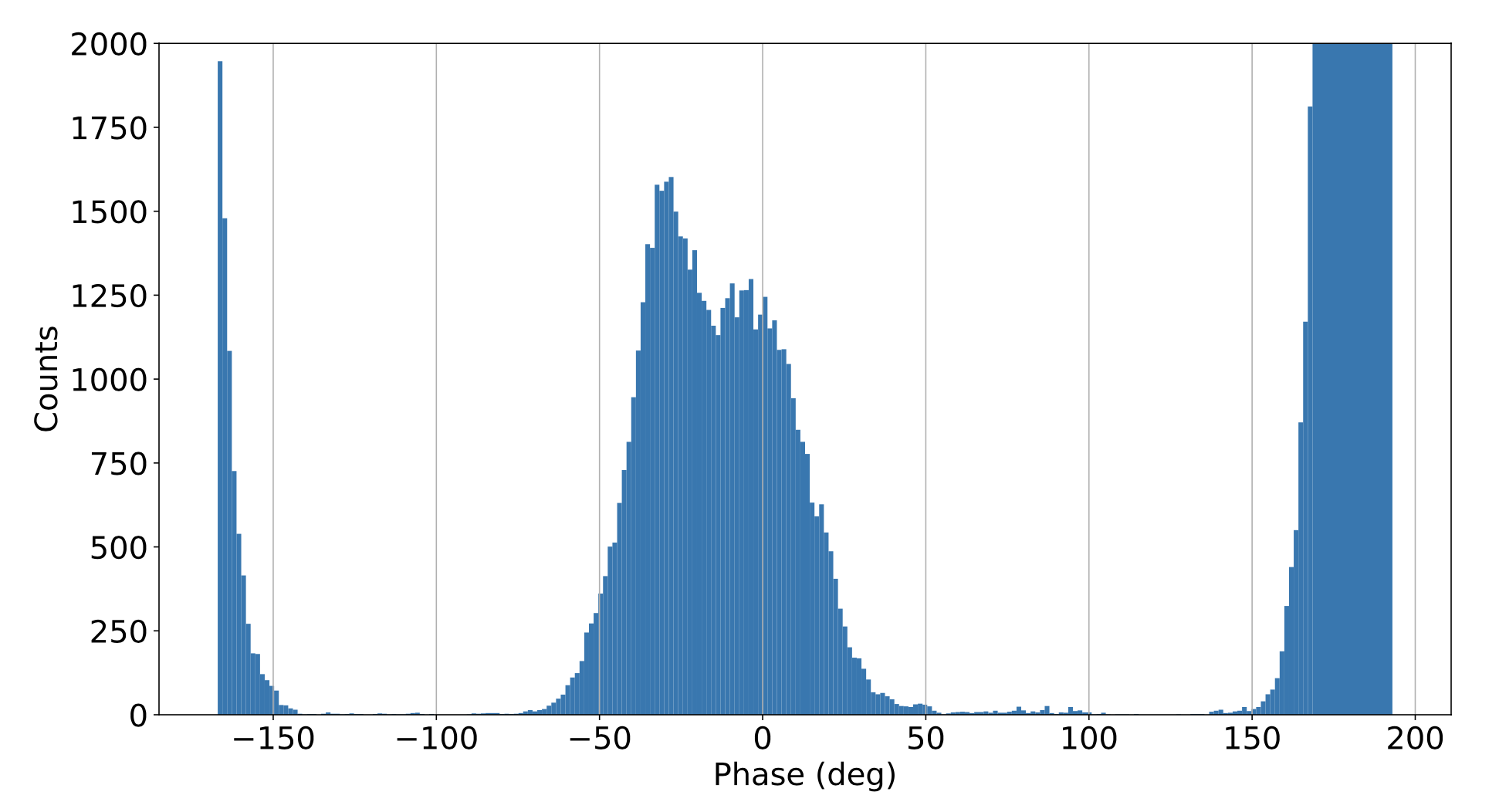}
    \caption{Histogram of the wavelet $CP-I$ phase measurements in the 3 minute band (1 min width), corresponding to large coherence ($>90$\%).}
    \label{fig:4}
\end{figure}

As mentioned in the previous section, the 3-minute band shows both intensity and CP oscillations. In order to possibly disentangle opacity effects from intrinsic magnetic oscillations, it is important to study the phase relations between the two diagnostics. While for long period CP oscillations there are no associated intensity oscillations, the examination of the phase lag is particularly important in the 3-minute band. Since both diagnostics display oscillations, it would be impossible to rule out opacity effects otherwise. To this aim, we consider the regions of the wavelet diagrams at which a high coherence is measured between the two time series. This ensures the reliability of the phase measurements themselves.
The light blue contours in Fig.~\ref{fig:2} highlight the regions of the wavelet diagrams associated with such highly coherent oscillations (coherence larger than $90\%$), which are subsequently considered in our analyses.
Since the wavelet analysis is performed on a pixel-by-pixel basis, we can study the occurrence of phase measurements in the 3-minute band, alongside their spatial distribution. 
The histogram of the wavelet phase measurements is shown in Fig.~\ref{fig:4}.  
Here, we note the presence of three components: one centred at $\pm 180$~degrees, and two partially overlapped but still distinguishable centred at $0$~degrees and $-35$~degrees. 
While the first two cannot exclude opacity effects as a potential cause, the latter is at a finite value and not consistent with opacity effects. Here it is worth noting that the phase values were already corrected for the effect of the sequential temporal scanning of the spectral line operated by IBIS (i.e. images at the different wavelengths are acquired sequentially and not strictly simultaneously).
Therefore, we interpret the phase shift at $-35$~degrees to be a physical phase shift, and statistically significant as a result of the number of occurrences demonstrated in Fig.~{\ref{fig:4}}. Finally, we note that the blue dots are found in the region where the measured CP signal is enhanced. This effect could be due to the density fluctuations associated to the waves, resulting in a shift of the geometrical height of formation of the spectral line.

\subsection{Magnetic Field Inclination}
\begin{figure}
    \centering
    \includegraphics[trim=10 0 0 0,clip,scale=0.25]{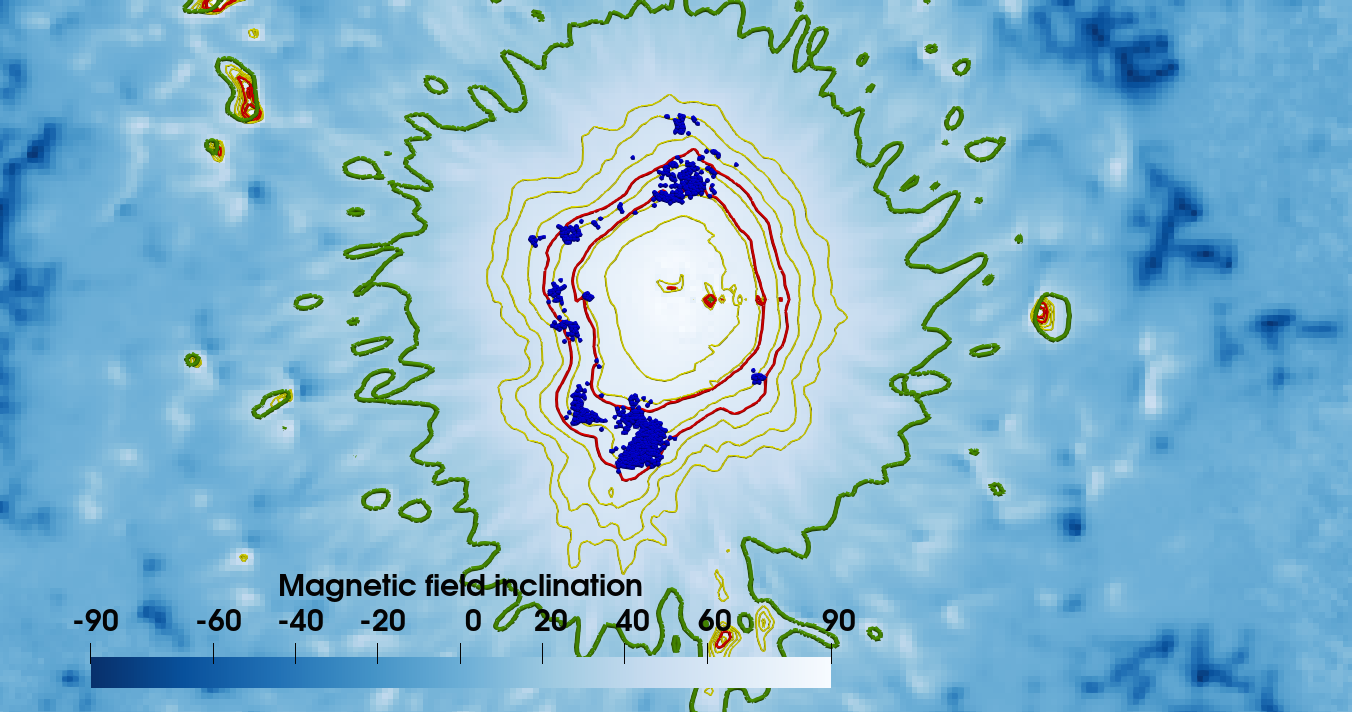}
    \caption{Photospheric magnetic field elevation angle with respect to the photospheric plane in the sunspot area, from the SHARP SDO/HMI data. The inclination is shown in blue shades, the yellow isolines are isocontours of the inclination of the field lines. The two red contours indicate the inclination between $63$ and $69$ degrees respectively. The green isolines show $B=500$~G (vertical magnetic field component), the blue dots are a regular subset of the blue dots in the central panel of Fig.~\ref{fig:1}.}
    \label{fig:5}
\end{figure}

More information regarding the properties of the phase lags can be gained through examination of the spatial distribution of these phase measurements in the FoV, which are indicated by the blue dots in Fig.~\ref{fig:1}. 
Interestingly, these blue dots are not spread across the FoV, but are instead localised approximately where the amplitude of the $CP$ oscillations are largest (see Fig.~\ref{fig:2}).
More importantly, the blue dots, which are the locations where both the intensity and $CP$ signals are linked by a statistically significant phase relation, appear to correspond to a particular narrow range of magnetic field inclinations spanning $63 \le \theta \le 69$~degrees.
This can be seen in Fig.~\ref{fig:5}, where we show the magnetic field inclination angles obtained from magnetograms captured by the Helioseismic and Magnetic Imager (HMI; \cite{2012SoPh..275..207S}) onboard the Solar Dynamics Observatory (SDO; \cite{2012SoPh..275....3P}). 
The alignment between the HMI magnetograms and the locations of the highly coherent  $-35$-degree phase lags established in the IBIS data was obtained through coalignment of the IBIS images with HMI continuum and Atmospheric Imaging Assembly (AIA; \cite{2012SoPh..275...17L}) EUV images (Fig.~{\ref{fig:1}}). 
It is remarkable that the clustering of the $\approx-35^\circ$ phase lag locations occur across a 6-degree span of magnetic field inclination angles. 
Such  a narrow range of magnetic field inclination values suggests a possible dependence of the $-35^\circ$ phase lag locations on the magnetic field geometry of the structure.

\section{Discussion and Conclusions}
The presence of observed magnetic oscillations associated with MHD waves is long debated (see for instance \citet{2015LRSP...12....6K}; and references therein).
In this work we have investigated the relation between CP and intensity oscillations and their spatial distribution in the chromosphere of a big sunspot observed by IBIS in the attempt to identify possible intrinsic magnetic oscillations.
As suggested by many authors in the past, one way to exclude the possibility that $CP$ oscillations are not due to cross-talk or opacity effects is to check for the phase lag with different quantities \cite{1999SoPh..187..389B, 2000SoPh..191...97K, 2009ApJ...702.1443F, 2015A&A...579A..73M}. 
We found that the umbra of the sunspot is dominated by 3-minute CP oscillations and these are associated to intensity oscillations. However, long period CP oscillations (up to 15 min) are also occasionally excited at the same locations, but in contrast to the 3-minute band, in this case we observe a lack of associated intensity oscillations. This is not consistent with opacity effects and suggests magnetic oscillations.\\
Regarding the 3-minute band, the simultaneous presence of intensity oscillations does not allow us to immediately rule out the possibility of opacity effects, therefore, a phase lag analysis between CP and intensity was performed. 
A phase relation between $CP$ and intensity is found at specific spatial locations in the sunspot. These phase measurements have a high confidence level, and they also appear not in agreement with opacity effects, thus suggesting the presence of intrinsic magnetic oscillations.

Let us now discuss the significance of a $-35$ degree phase lag. There are two main factors that can contribute to the measured phase component that appears to be related to wave activity; intrinsic phase lag of the wave mode and the height difference at which the CP and intensity originate. It has to be noted that these two are not mutually exclusive and may both come into play at the same time resulting in the final measured phase lag. 
In other words, the resulting measured phase $\phi$ can be thought of as the combination of an intrinsic wave term $\phi_{wave}$ and a second contribution $\phi_{\Delta h}$ given by the height difference of the CP and intensity signals in the case of a vertically propagating wave.
Regarding the first contribution, as mentioned in the introduction, different wave modes can have different phase relations between their diagnostics. Indeed, this can in principle be exploited to identify the modes themselves \citep{2009ApJ...702.1443F, 2013A&A...551A.137M}. However, this is not straightforward in the case of complex geometries where a specific propagation state or wave reflection can hamper the correct wave identification.  A significant improvement could be represented by the use of realistic state-of-the-art numerical simulations, where specific modes are excited in complex geometries to estimate the phase lag between different types of fluctuations (e.g. phase lag between velocity and magnetic field fluctuations), which could be then used to interpret observational results such as ours.\\
The second contribution to the phase lag is related to the intrinsic difference between the height where the CP signal and the core intensity signal originate. Such a difference can introduce an additional phase term in the case of a propagating wave. The problem of the determination of the height of formation of the intensity and magnetic signals in the Ca $854.2$ nm spectral line was addressed in great detail in \citet{Cauzzi2008, 2017MNRAS.464.4534Q, 2019ApJ...873..126M}. In addition, the height difference can also dynamically change in a magneto-hydrodynamic atmosphere, a circumstance that may also contribute to the width of the distribution of the phase measurements seen in Fig. \ref{fig:4}. 
Although we cannot completely rule out this second possibility, in the case of an intense magnetic field strength such as the one considered here (i.e. $4$ kG in the photosphere as inferred from Hinode level 2 inverted data), we expect the Alfv\'en speed to be so large ($>50$ km/s) that the contribution to the measured phase lag from $\phi_{wave}$ is expected to be significantly large than the contribution from $\phi_{\Delta h}$. 
Indeed, given the fact that the time difference corresponding to the measured phase lag at a period of 3 min ($5$ mHz) is $\sim 20$ s, the height difference that would cause such a phase delay would be of the order of at least  $1000$ km, which is larger than expected from the results presented in the articles mentioned above. Finally, we recall that along with the $-35$ degrees component, two additional components are detected at $0$ and $\pm \pi$, which are likely due to opacity effects. If the height difference contribution were to be significant, then these additional components should also be displaced along the phase axis, which is not the case. 
Therefore, we conclude that, regardless of the exact nature of the contributions to the phase lag discussed above, the mere existence of a non-opacity-related contribution (the $-35$ degree component in the distribution) corroborates the presence of magnetic-like waves (i.e. fast magneto-acoustic mode in the low-$\beta$ regime, being clearly distinguishable from the other two components).
Our interpretation is also in agreement with the results of \citet{2020ApJ...892...49H}, who have found Alfv{\'{e}}n/intermediate shocks at the same spatial locations. In this regard, given the fact that the coherence patches in the wavelet diagrams can extend for a few periods, we believe that our results may be interpreted as the signature of the magneto-acoustic fast mode that eventually leads to impulsive Alfvénic shocks which are followed by sudden reversals in the Stokes profiles.

\citet{Baker2020} have found by using \emph{Hinode}/EIS observations that the same active region shows the presence of elemental abundance anomalies on the same side of the umbra. 
As previously mentioned, the detection of magnetic fluctuations associated with MHD waves takes on a particular importance for the FIP effect \cite{2015LRSP...12....2L}. 
In this regard, the observations combined with the Laming fractionation model suggest a possible link between the co-spatial CP perturbations and the anomalous abundances.\\
It is worth highlighting that the narrow range of inclination angles of the magnetic field, corresponding to the locations at which a phase relation between intensity and $CP$ perturbations are found, suggests a possible role of the magnetic field inclination and geometry. It has been shown \cite{2006MNRAS.372..551S, 2006ApJ...653..739K, 2008SoPh..251..251C, 2010ApJ...719..357F} that at the Alfv{\'{e}}n-acoustic equipartition, magneto-acoustic waves can be converted into other modes (e.g. from an acoustic-like mode to a magnetic-like mode as the one we observe here and viceversa).
Furthermore, it has been shown \cite{2006MNRAS.372..551S} that the efficiency with which modes are converted (i.e. the amount of energy contained in a single mode which is converted into another mode) depends on the angle between the wavevector and the magnetic field lines.
For this reason, one possibility could be that the magnetic disturbances observed in this work might result from the mode conversion of fast (acoustic-like) magneto-acoustic waves (in the high plasma-$\beta$ regime) to a combination of fast (magnetic-like) magneto-acoustic waves (in a low plasma-$\beta$ regime synonymous with sunspot chromospheres  \cite{2001SoPh..203...71G, 2011ASInC...2..221C}). This interpretation agrees with the findings put forward by \citet{2020ApJ...892...49H}, who calculated the plasma-$\beta=1$ isocontours at photospheric and chromospheric heights for the same IBIS dataset.
Indeed, it was found that vertically propagating waves, in the region where the Alfvén shocks were observed, encounter the equipartition layer before reaching the chromosphere, therefore suggesting a possible mode conversion as the cause of the magnetic fluctuations.
While the narrow range of inclinations associated to the detected magnetic fluctuations in our study may suggest a possible role of the mode conversion here, it is important to note that if the mode conversion was the only cause, we would have expected a symmetric distribution of the magnetic waves themselves. But the blue dots are only found on the left side of the sunspot. 
However, from AIA observations (see Fig. \ref{fig:1}), we argue that, although in the lower atmosphere the sunspot appears quite symmetric, higher up its connectivity with surrounding field appears not to be so. In this regard, \citet{2015LRSP...12....2L} have already suggested that magnetic waves responsible for the FIP effect may be generated at coronal heights by e.g. nanoflare activity. Therefore, the magnetic field connectivity may also play a role and our results appear to suggest that this is the case.
In other words, the spatial asymmetry of the distribution of magnetic fluctuations highlights the need for a more detailed analysis of the effect global magnetic geometries have on the transmission and conversion of various MHD wave modes in sunspot atmospheres, which is beyond the scope of our current study. We anticipate follow-up work employing spectropolarimetric inversions to further understand the role of magnetic field geometries in chromospheric MHD wave physics. 
Finally, we would like to stress once more that understanding the mechanisms that generate the FIP effect can be helpful to establish a link between the solar atmosphere and the heliosphere. In this regard, Solar Orbiter will enable a more detailed analysis of the mechanisms underlying the FIP effect itself, and the study and identification of magnetic-like waves in the solar atmosphere can provide additional information to be incorporated into models to further advance our knowledge.

\enlargethispage{20pt}

\dataccess{The IBIS data that support the plots within this paper and other findings of this study can be downloaded from the IBIS-A archive \href{http://ibis.oa-roma.inaf.it/IBISA/database/}{http://ibis.oa-roma.inaf.it/IBISA/database/}.}

\aucontribute{MS carried out the experiments and conceived of and designed the study. MS performed the data reduction and scientific analysis, with assistance from DB, GV, DBJ, SJ, MM, ASHT, DHB, IE, FG, and CDM. MS drafted the manuscript with the help of DB, GV, DJ and SJ. All authors read and approved the manuscript.}

\competing{The author(s) declare that they have no competing interests.}

\funding{This work was supported by: 
\vspace{-4mm}
\begin{itemize}
    \item An Invest NI and Randox Laboratories Ltd. Research \& Development Grant (059RDEN-1);
    \item European Research Council (grant agreement No. 682462);
    \item Research Council of Norway (project No. 262622);
    \item The Department for the Economy (Northern Ireland) through their postgraduate research studentship. 
    \item European Union’s Horizon 2020 Research and Innovation program under grant agreement No 82135 (SOLARNET) and No 739500 (PRE-EST)
    \end{itemize}}

\ack{The authors wish to acknowledge scientific discussions with the Waves in the Lower Solar Atmosphere (WaLSA; \href{https://www.WaLSA.team}{www.WaLSA.team}) team, which is supported by the Research Council of Norway (project number 262622), and The Royal Society through the award of funding to host the Theo Murphy Discussion Meeting “High resolution wave dynamics in the lower solar atmosphere” (grant Hooke18b/SCTM). DBJ is grateful to Invest NI and Randox Laboratories Ltd. for the award of a Research \& Development Grant (059RDEN-1). SJ acknowledges support from the European Research Council under the European Unions Horizon 2020 research and innovation program (grant agreement No. 682462) and from the Research Council of Norway through sits Centres of Excellence scheme (project No. 262622). CDM would like to thank the Northern Ireland Department for the Economy for the award of a PhD studentship. D.B. is funded under STFC consolidated grant No. ST/S000240/1.
This research has received funding from the European Union’s Horizon 2020 Research and Innovation program under grant agreement No 824135 (SOLARNET), the  STFC grant number ST/T000317/1, and No 739500 (PRE-EST). This research has made use of the IBIS-A archive.
 }


\bibliographystyle{rstasj}

\begin{thebibliography}{79}
\providecommand{\natexlab}[1]{#1}
\providecommand{\url}[1]{\texttt{#1}}
\expandafter\ifx\csname urlstyle\endcsname\relax
  \providecommand{\doi}[1]{doi: #1}\else
  \providecommand{\doi}{doi: \begingroup \urlstyle{rm}\Url}\fi

\bibitem[Edwin and Roberts(1983)]{Edwin1983}
Edwin PM, Roberts B.
\newblock 1983, {Wave propagation in a magnetic cylinder}.
\newblock \emph{Sol. Phys.}, 88\penalty0 (1-2):\penalty0 179--191.
\newblock ISSN 0038-0938.
\newblock (\doi{10.1007/BF00196186}).
\newblock URL \url{http://adsabs.harvard.edu/abs/1983SoPh...88..179E}

\bibitem[Roberts(1983)]{Roberts1983}
Roberts B.
\newblock 1983, {Wave propagation in intense flux tubes}.
\newblock \emph{Sol. Phys.}, 87\penalty0 (1):\penalty0 77--93.
\newblock ISSN 0038-0938.
\newblock (\doi{10.1007/BF00151162}).
\newblock URL \url{http://adsabs.harvard.edu/abs/1983SoPh...87...77R}

\bibitem[Khomenko et~al.(2008)Khomenko, Collados, and Felipe]{Khomenko2008}
Khomenko E, Collados M, Felipe T.
\newblock 2008, {Nonlinear Numerical Simulations of Magneto-Acoustic Wave
  Propagation in Small-Scale Flux Tubes}.
\newblock \emph{Sol. Phys.}, 251\penalty0 (1-2):\penalty0 589--611.
\newblock ISSN 0038-0938.
\newblock (\doi{10.1007/s11207-008-9133-8}).
\newblock URL \url{http://adsabs.harvard.edu/abs/2008SoPh..251..589K}

\bibitem[{Felipe} et~al.(2010){Felipe}, {Khomenko}, and
  {Collados}]{2010ApJ...719..357F}
{Felipe} T, {Khomenko} E, {Collados} M.
\newblock 2010, {Magneto-acoustic Waves in Sunspots: First Results From a New
  Three-dimensional Nonlinear Magnetohydrodynamic Code}.
\newblock \emph{ApJ}, 719\penalty0 (1):\penalty0 357--377.
\newblock (\doi{10.1088/0004-637X/719/1/357})

\bibitem[Fedun et~al.(2011)Fedun, Shelyag, and
  Erd{\'{e}}lyi]{2011ApJ...727...17F}
Fedun V, Shelyag S, Erd{\'{e}}lyi R.
\newblock 2011, {Numerical Modeling of Footpoint-driven Magneto-acoustic Wave
  Propagation in a Localized Solar Flux Tube}.
\newblock \emph{ApJ}, 727:\penalty0 17.
\newblock (\doi{10.1088/0004-637X/727/1/17})

\bibitem[Mumford and Erd{\'{e}}lyi(2015)]{2015MNRAS.449.1679M}
Mumford SJ, Erd{\'{e}}lyi R.
\newblock 2015, {Photospheric logarithmic velocity spirals as MHD wave
  generation mechanisms}.
\newblock \emph{MNRAS}, 449:\penalty0 1679--1685.
\newblock (\doi{10.1093/mnras/stv365})

\bibitem[Mumford et~al.(2015)Mumford, Fedun, and
  Erd{\'{e}}lyi]{2015ApJ...799....6M}
Mumford SJ, Fedun V, Erd{\'{e}}lyi R.
\newblock 2015, {Generation of Magnetohydrodynamic Waves in Low Solar
  Atmospheric Flux Tubes by Photospheric Motions}.
\newblock \emph{ApJ}, 799:\penalty0 6.
\newblock (\doi{10.1088/0004-637X/799/1/6})

\bibitem[L{\"{o}}hner-B{\"{o}}ttcher(2016)]{2016PhDT........15L}
L{\"{o}}hner-B{\"{o}}ttcher J.
\newblock 2016, \emph{{Wave phenomena in sunspots}}.
\newblock PhD thesis, Universit{\"{a}}t Freiburg im Breisgau

\bibitem[Grant et~al.(2018)Grant, Jess, Zaqarashvili, Beck, Socas-Navarro,
  Aschwanden, Keys, Christian, Houston, and Hewitt]{Grant:2018vfa}
Grant SDT, Jess DB, Zaqarashvili TV, Beck C, Socas-Navarro H, Aschwanden MJ,
  Keys PH, Christian DJ, Houston SJ, Hewitt RL.
\newblock 2018, {Alfv{\'{e}}n wave dissipation in the solar chromosphere}.
\newblock \emph{Nature Physics}, 14\penalty0 (5):\penalty0 480--483

\bibitem[Keys et~al.(2018)Keys, Morton, Jess, Verth, Grant, Mathioudakis,
  Mackay, Doyle, Christian, Keenan, and Erd{\'{e}}lyi]{2018ApJ...857...28K}
Keys PH, Morton RJ, Jess DB, Verth G, Grant SDT, Mathioudakis M, Mackay DH,
  Doyle JG, Christian DJ, Keenan FP, Erd{\'{e}}lyi R.
\newblock 2018, {Photospheric Observations of Surface and Body Modes in Solar
  Magnetic Pores}.
\newblock \emph{ApJ}, 857:\penalty0 28.
\newblock (\doi{10.3847/1538-4357/aab432})

\bibitem[Centeno et~al.(2006)Centeno, Collados, and
  Trujillo~Bueno]{2006ApJ...640.1153C}
Centeno R, Collados M, Trujillo~Bueno J.
\newblock 2006, {Spectropolarimetric Investigation of the Propagation of
  Magnetoacoustic Waves and Shock Formation in Sunspot Atmospheres}.
\newblock \emph{ApJ}, 640:\penalty0 1153--1162.
\newblock (\doi{10.1086/500185})

\bibitem[Jess et~al.(2009)Jess, Mathioudakis, Erd{\'{e}}lyi, Crockett, Keenan,
  and Christian]{2009Jess}
Jess DB, Mathioudakis M, Erd{\'{e}}lyi R, Crockett PJ, Keenan FP, Christian DJ.
\newblock 2009, {Alfv{\'{e}}n Waves in the Lower Solar Atmosphere}.
\newblock \emph{Science}, 323:\penalty0 1582--.
\newblock (\doi{10.1126/science.1168680})

\bibitem[Felipe et~al.(2011)Felipe, Khomenko, and
  Collados]{2011ApJ...735...65F}
Felipe T, Khomenko E, Collados M.
\newblock 2011, {Magnetoacoustic Wave Energy from Numerical Simulations of an
  Observed Sunspot Umbra}.
\newblock \emph{ApJ}, 735:\penalty0 65.
\newblock (\doi{10.1088/0004-637X/735/1/65})

\bibitem[Morton et~al.(2012)Morton, Verth, Jess, Kuridze, Ruderman,
  Mathioudakis, and Erd{\'{e}}lyi]{2012NatCo...3E1315M}
Morton RJ, Verth G, Jess DB, Kuridze D, Ruderman MS, Mathioudakis M,
  Erd{\'{e}}lyi R.
\newblock 2012, {Observations of ubiquitous compressive waves in the Sun's
  chromosphere}.
\newblock \emph{Nature Communications}, 3:\penalty0 1315.
\newblock (\doi{10.1038/ncomms2324})

\bibitem[Mathioudakis et~al.(2013)Mathioudakis, Jess, and
  Erd{\'{e}}lyi]{2013SSRv..175....1M}
Mathioudakis M, Jess DB, Erd{\'{e}}lyi R.
\newblock 2013, {Alfv{\'{e}}n Waves in the Solar Atmosphere. From Theory to
  Observations}.
\newblock \emph{Space Sci. Rev.}, 175:\penalty0 1--27.
\newblock (\doi{10.1007/s11214-012-9944-7})

\bibitem[Jess et~al.(2015)Jess, Morton, Verth, Fedun, Grant, and
  Giagkiozis]{2015SSRv..190..103J}
Jess DB, Morton RJ, Verth G, Fedun V, Grant SDT, Giagkiozis I.
\newblock 2015, {Multiwavelength Studies of MHD Waves in the Solar
  Chromosphere. An Overview of Recent Results}.
\newblock \emph{Space Sci. Rev.}, 190:\penalty0 103--161.
\newblock (\doi{10.1007/s11214-015-0141-3})

\bibitem[Krishna~Prasad et~al.(2017)Krishna~Prasad, Jess, Van~Doorsselaere,
  Verth, Morton, Fedun, Erd{\'{e}}lyi, and Christian]{2017ApJ...847....5K}
Krishna~Prasad S, Jess DB, Van~Doorsselaere T, Verth G, Morton RJ, Fedun V,
  Erd{\'{e}}lyi R, Christian DJ.
\newblock 2017, {The Frequency-dependent Damping of Slow Magnetoacoustic Waves
  in a Sunspot Umbral Atmosphere}.
\newblock \emph{ApJ}, 847:\penalty0 5.
\newblock (\doi{10.3847/1538-4357/aa86b5})

\bibitem[McIntosh and Jefferies(2006)]{2006ApJ...647L..77M}
McIntosh SW, Jefferies SM.
\newblock 2006, {Observing the Modification of the Acoustic Cutoff Frequency by
  Field Inclination Angle}.
\newblock \emph{ApJl}, 647:\penalty0 L77--L81.
\newblock (\doi{10.1086/507425})

\bibitem[Jefferies et~al.(2006)Jefferies, McIntosh, Armstrong, Bogdan,
  Cacciani, and Fleck]{2006ApJ...648L.151J}
Jefferies SM, McIntosh SW, Armstrong JD, Bogdan TJ, Cacciani A, Fleck B.
\newblock 2006, {Magnetoacoustic Portals and the Basal Heating of the Solar
  Chromosphere}.
\newblock \emph{ApJl}, 648:\penalty0 L151--L155.
\newblock (\doi{10.1086/508165})

\bibitem[Bloomfield et~al.(2007)Bloomfield, Lagg, and
  Solanki]{2007ApJ...671.1005B}
Bloomfield DS, Lagg A, Solanki SK.
\newblock 2007, {The Nature of Running Penumbral Waves Revealed}.
\newblock \emph{ApJ}, 671:\penalty0 1005--1012.
\newblock (\doi{10.1086/523266})

\bibitem[{Stangalini} et~al.(2011){Stangalini}, {Del Moro}, {Berrilli}, and
  {Jefferies}]{2011A&A...534A..65S}
{Stangalini} M, {Del Moro} D, {Berrilli} F, {Jefferies} SM.
\newblock 2011, {MHD wave transmission in the Sun's atmosphere}.
\newblock \emph{A\&A}, 534:\penalty0 A65.
\newblock (\doi{10.1051/0004-6361/201117356})

\bibitem[{Moreels} et~al.(2015){Moreels}, {Freij}, {Erd{\'e}lyi}, {Van
  Doorsselaere}, and {Verth}]{2015A&A...579A..73M}
{Moreels} MG, {Freij} N, {Erd{\'e}lyi} R, {Van Doorsselaere} T, {Verth} G.
\newblock 2015, {Observations and mode identification of sausage waves in a
  magnetic pore}.
\newblock \emph{A\&A}, 579:\penalty0 A73.
\newblock (\doi{10.1051/0004-6361/201425096})

\bibitem[Allcock and Erd{\'{e}}lyi(2017)]{2017SoPh..292...35A}
Allcock M, Erd{\'{e}}lyi R.
\newblock 2017, {Magnetohydrodynamic Waves in an Asymmetric Magnetic Slab}.
\newblock \emph{Sol. Phys.}, 292:\penalty0 35.
\newblock (\doi{10.1007/s11207-017-1054-y})

\bibitem[Zs{\'{a}}mberger et~al.(2018)Zs{\'{a}}mberger, Allcock, and
  Erd{\'{e}}lyi]{2018ApJ...853..136K}
Zs{\'{a}}mberger NK, Allcock M, Erd{\'{e}}lyi R.
\newblock 2018, {Magneto-acoustic Waves in a Magnetic Slab Embedded in an
  Asymmetric Magnetic Environment: The Effects of Asymmetry}.
\newblock \emph{ApJ}, 853:\penalty0 136.
\newblock (\doi{10.3847/1538-4357/aa9ffe})

\bibitem[Bogdan(2000)]{2000SoPh..192..373B}
Bogdan TJ.
\newblock 2000, {Sunspot Oscillations: A Review - (Invited Review)}.
\newblock \emph{Sol. Phys.}, 192:\penalty0 373--394.
\newblock (\doi{10.1023/A:1005225214520})

\bibitem[{Chorley} et~al.(2010){Chorley}, {Hnat}, {Nakariakov}, {Inglis}, and
  {Bakunina}]{2010A&A...513A..27C}
{Chorley} N, {Hnat} B, {Nakariakov} VM, {Inglis} AR, {Bakunina} IA.
\newblock 2010, {Long period oscillations in sunspots}.
\newblock \emph{A\&A}, 513:\penalty0 A27.
\newblock (\doi{10.1051/0004-6361/200913683})

\bibitem[Morton et~al.(2011)Morton, Erd{\'{e}}lyi, Jess, and
  Mathioudakis]{2011ApJ...729L..18M}
Morton RJ, Erd{\'{e}}lyi R, Jess DB, Mathioudakis M.
\newblock 2011, {Observations of Sausage Modes in Magnetic Pores}.
\newblock \emph{ApJl}, 729:\penalty0 L18.
\newblock (\doi{10.1088/2041-8205/729/2/L18})

\bibitem[{Stangalini} et~al.(2012){Stangalini}, {Giannattasio}, {Del Moro}, and
  {Berrilli}]{2012A&A...539L...4S}
{Stangalini} M, {Giannattasio} F, {Del Moro} D, {Berrilli} F.
\newblock 2012, {Three-minute wave enhancement in the solar photosphere}.
\newblock \emph{A\&A}, 539:\penalty0 L4.
\newblock (\doi{10.1051/0004-6361/201118654})

\bibitem[Grant et~al.(2015)Grant, Jess, Moreels, Morton, Christian, Giagkiozis,
  Verth, Fedun, Keys, Van~Doorsselaere, and Erd{\'{e}}lyi]{2015ApJ...806..132G}
Grant SDT, Jess DB, Moreels MG, Morton RJ, Christian DJ, Giagkiozis I, Verth G,
  Fedun V, Keys PH, Van~Doorsselaere T, Erd{\'{e}}lyi R.
\newblock 2015, {Wave Damping Observed in Upwardly Propagating Sausage-mode
  Oscillations Contained within a Magnetic Pore}.
\newblock \emph{ApJ}, 806:\penalty0 132.
\newblock (\doi{10.1088/0004-637X/806/1/132})

\bibitem[Jafarzadeh et~al.(2017)Jafarzadeh, Solanki, Stangalini, Steiner,
  Cameron, and Danilovic]{2017ApJS..229...10J}
Jafarzadeh S, Solanki SK, Stangalini M, Steiner O, Cameron RH, Danilovic S.
\newblock 2017, {High-frequency Oscillations in Small Magnetic Elements
  Observed with Sunrise/SuFI}.
\newblock \emph{ApJs}, 229:\penalty0 10.
\newblock (\doi{10.3847/1538-4365/229/1/10})

\bibitem[Jess et~al.(2017)Jess, Van~Doorsselaere, Verth, Fedun, Krishna~Prasad,
  Erd{\'{e}}lyi, Keys, Grant, Uitenbroek, and Christian]{2017ApJ...842...59J}
Jess DB, Van~Doorsselaere T, Verth G, Fedun V, Krishna~Prasad S, Erd{\'{e}}lyi
  R, Keys PH, Grant SDT, Uitenbroek H, Christian DJ.
\newblock 2017, {An Inside Look at Sunspot Oscillations with Higher Azimuthal
  Wavenumbers}.
\newblock \emph{ApJ}, 842:\penalty0 59.
\newblock (\doi{10.3847/1538-4357/aa73d6})

\bibitem[Fujimura and Tsuneta(2009)]{2009ApJ...702.1443F}
Fujimura D, Tsuneta S.
\newblock 2009, {Properties of Magnetohydrodynamic Waves in the Solar
  Photosphere Obtained with Hinode}.
\newblock \emph{ApJ}, 702:\penalty0 1443--1457.
\newblock (\doi{10.1088/0004-637X/702/2/1443})

\bibitem[{Moreels, M. G.} and {Van Doorsselaere, T.}(2013)]{Moreels2013}
{Moreels, M. G.}, {Van Doorsselaere, T.}
\newblock 2013, Phase relations for seismology of photospheric flux tubes.
\newblock \emph{A\&A}, 551:\penalty0 A137.
\newblock (\doi{10.1051/0004-6361/201219568}).
\newblock URL \url{https://doi.org/10.1051/0004-6361/201219568}

\bibitem[{Moreels, M. G.} et~al.(2013){Moreels, M. G.}, {Goossens, M.}, and
  {Van Doorsselaere, T.}]{MoreelsGoossens}
{Moreels, M. G.}, {Goossens, M.}, {Van Doorsselaere, T.}
\newblock 2013, Cross-sectional area and intensity variations of sausage modes.
\newblock \emph{A\&A}, 555:\penalty0 A75.
\newblock (\doi{10.1051/0004-6361/201321545}).
\newblock URL \url{https://doi.org/10.1051/0004-6361/201321545}

\bibitem[Team et~al.(2019)Team, Al-Janabi, Antolin, Baker, Bellot~Rubio,
  Bradley, Brooks, Centeno, Culhane, Del~Zanna, Doschek, Fletcher, Hara, Harra,
  Hillier, Imada, Klimchuk, Mariska, Pereira, Reeves, Sakao, Sakurai, Shimizu,
  Shimojo, Shiota, Solanki, Sterling, Su, Suematsu, Tarbell, Tiwari, Toriumi,
  Ugarte-Urra, Warren, Watanabe, and Young]{10.1093/pasj/psz084}
Team HR, Al-Janabi K, Antolin P, Baker D, Bellot~Rubio LR, Bradley L, Brooks
  DH, Centeno R, Culhane JL, Del~Zanna G, Doschek GA, Fletcher L, Hara H, Harra
  LK, Hillier AS, Imada S, Klimchuk JA, Mariska JT, Pereira TMD, Reeves KK,
  Sakao T, Sakurai T, Shimizu T, Shimojo M, Shiota D, Solanki SK, Sterling AC,
  Su Y, Suematsu Y, Tarbell TD, Tiwari SK, Toriumi S, Ugarte-Urra I, Warren HP,
  Watanabe T, Young PR.
\newblock 2019, {Achievements of Hinode in the first eleven years}.
\newblock \emph{Publications of the Astronomical Society of Japan}, 71\penalty0
  (5).
\newblock ISSN 0004-6264.
\newblock (\doi{10.1093/pasj/psz084}).
\newblock URL \url{https://doi.org/10.1093/pasj/psz084}.
\newblock R1

\bibitem[{Laming}(2015)]{2015LRSP...12....2L}
{Laming} JM.
\newblock 2015, {The FIP and Inverse FIP Effects in Solar and Stellar Coronae}.
\newblock \emph{Living Reviews in Solar Physics}, 12\penalty0 (1):\penalty0 2.
\newblock (\doi{10.1007/lrsp-2015-2})

\bibitem[{M{\"u}ller} et~al.(2013){M{\"u}ller}, {Marsden}, {St. Cyr}, and
  {Gilbert}]{2013SoPh..285...25M}
{M{\"u}ller} D, {Marsden} RG, {St. Cyr} OC, {Gilbert} HR.
\newblock 2013, {Solar Orbiter . Exploring the Sun-Heliosphere Connection}.
\newblock \emph{Sol Phys}, 285\penalty0 (1-2):\penalty0 25--70.
\newblock (\doi{10.1007/s11207-012-0085-7})

\bibitem[{Laming} et~al.(2019){Laming}, {Vourlidas}, {Korendyke}, {Chua},
  {Cranmer}, {Ko}, {Kuroda}, {Provornikova}, {Raymond}, {Raouafi}, {Strachan},
  {Tun-Beltran}, {Weberg}, and {Wood}]{2019ApJ...879..124L}
{Laming} JM, {Vourlidas} A, {Korendyke} C, {Chua} D, {Cranmer} SR, {Ko} YK,
  {Kuroda} N, {Provornikova} E, {Raymond} JC, {Raouafi} NE, {Strachan} L,
  {Tun-Beltran} S, {Weberg} M, {Wood} BE.
\newblock 2019, {Element Abundances: A New Diagnostic for the Solar Wind}.
\newblock \emph{ApJ}, 879\penalty0 (2):\penalty0 124.
\newblock (\doi{10.3847/1538-4357/ab23f1})

\bibitem[Horn et~al.(1997)Horn, Staude, and Landgraf]{1997SoPh..172...69H}
Horn T, Staude J, Landgraf V.
\newblock 1997, {Observations of Sunspot Umbral Oscillations}.
\newblock \emph{Sol. Phys.}, 172:\penalty0 69--76.
\newblock (\doi{10.1023/A:1004909030878})

\bibitem[Norton et~al.(1999)Norton, Ulrich, Bush, and Tarbell]{Norton_1999}
Norton AA, Ulrich RK, Bush RI, Tarbell TD.
\newblock 1999, Characteristics of magnetohydrodynamic oscillations observed
  with the michelson doppler imager.
\newblock \emph{The Astrophysical Journal}, 518\penalty0 (2):\penalty0
  L123--L126.
\newblock (\doi{10.1086/312072}).
\newblock URL \url{https://doi.org/10.1086%2F312072}

\bibitem[R{\"{u}}edi et~al.(1999)R{\"{u}}edi, Solanki, Bogdan, and
  Cally]{1999ASSL..243..337R}
R{\"{u}}edi I, Solanki SK, Bogdan T, Cally P.
\newblock 1999,
\newblock , \emph{Polarization}, 337--347volume 243 of \emph{Astrophysics and
  Space Science Library}pages.
\newblock (\doi{10.1007/978-94-015-9329-8{\_}29})

\bibitem[Balthasar(1999)]{1999SoPh..187..389B}
Balthasar H.
\newblock 1999, {Temporal fluctuations of the magnetic field in sunspots}.
\newblock \emph{Sol. Phys.}, 187:\penalty0 389--403.
\newblock (\doi{10.1023/A:1005131927915})

\bibitem[Staude(2002)]{2002AN....323..317S}
Staude J.
\newblock 2002, {Magnetic field oscillations of sunspots?}
\newblock \emph{Astronomische Nachrichten}, 323:\penalty0 317--320.
\newblock (\doi{10.1002/1521-3994(200208)323:3/4<317::AID-ASNA317>3.0.CO;2-Y})

\bibitem[Bellot~Rubio et~al.(2000)Bellot~Rubio, Collados, Ruiz~Cobo, and
  Rodr{\textbackslash}'{\textbackslash}iguez~Hidalgo]{2000ApJ...534..989B}
Bellot~Rubio LR, Collados M, Ruiz~Cobo B,
  Rodr{\textbackslash}'{\textbackslash}iguez~Hidalgo I.
\newblock 2000, {Oscillations in the Photosphere of a Sunspot Umbra from the
  Inversion of Infrared Stokes Profiles}.
\newblock \emph{ApJ}, 534:\penalty0 989--996.
\newblock (\doi{10.1086/308791})

\bibitem[{Rueedi} et~al.(1998){Rueedi}, {Solanki}, {Stenflo}, {Tarbell}, and
  {Scherrer}]{1998A&A...335L..97R}
{Rueedi} I, {Solanki} SK, {Stenflo} JO, {Tarbell} T, {Scherrer} PH.
\newblock 1998, {Oscillations of sunspot magnetic fields}.
\newblock \emph{A\&A}, 335:\penalty0 L97--L100

\bibitem[Kupke et~al.(2000)Kupke, Labonte, and Mickey]{2000SoPh..191...97K}
Kupke R, Labonte BJ, Mickey DL.
\newblock 2000, {Observational Study of Sunspot Oscillations in Stokes I, Q, U,
  and V}.
\newblock \emph{Sol. Phys.}, 191:\penalty0 97--128.
\newblock (\doi{10.1023/A:1005286619634})

\bibitem[{Gri{\~n}{\'o}n-Mar{\'\i}n} et~al.(2020){Gri{\~n}{\'o}n-Mar{\'\i}n},
  {Pastor Yabar}, {Socas-Navarro}, and {Centeno}]{2020A&A...635A..64G}
{Gri{\~n}{\'o}n-Mar{\'\i}n} AB, {Pastor Yabar} A, {Socas-Navarro} H, {Centeno}
  R.
\newblock 2020, {Discovery of long-period magnetic field oscillations and
  motions in isolated sunspots}.
\newblock (\doi{10.1051/0004-6361/201936589})

\bibitem[Lites et~al.(1998)Lites, Thomas, Bogdan, and
  Cally]{1998ApJ...497..464L}
Lites BW, Thomas JH, Bogdan TJ, Cally PS.
\newblock 1998, {Velocity and Magnetic Field Fluctuations in the Photosphere of
  a Sunspot}.
\newblock \emph{ApJ}, 497:\penalty0 464--482.
\newblock (\doi{10.1086/305451})

\bibitem[{Khomenko} and {Collados}(2015)]{2015LRSP...12....6K}
{Khomenko} E, {Collados} M.
\newblock 2015, {Oscillations and Waves in Sunspots}.
\newblock \emph{Living Reviews in Solar Physics}, 12\penalty0 (1):\penalty0 6.
\newblock (\doi{10.1007/lrsp-2015-6})

\bibitem[{R{\"u}edi} and {Cally}(2003)]{2003A&A...410.1023R}
{R{\"u}edi} I, {Cally} PS.
\newblock 2003, {A comparison between model calculations and observations of
  sunspot oscillations}.
\newblock \emph{A\&A}, 410:\penalty0 1023--1028.
\newblock (\doi{10.1051/0004-6361:20031331})

\bibitem[Mart{\textbackslash}'{\textbackslash}inez~Gonz{\'{a}}lez
  et~al.(2011)Mart{\textbackslash}'{\textbackslash}inez~Gonz{\'{a}}lez,
  Asensio~Ramos, Manso~Sainz, Khomenko,
  Mart{\textbackslash}'{\textbackslash}inez~Pillet, Solanki,
  L{\'{o}}pez~Ariste, Schmidt, Barthol, and Gandorfer]{2011ApJ...730L..37M}
Mart{\textbackslash}'{\textbackslash}inez~Gonz{\'{a}}lez MJ, Asensio~Ramos A,
  Manso~Sainz R, Khomenko E, Mart{\textbackslash}'{\textbackslash}inez~Pillet
  V, Solanki SK, L{\'{o}}pez~Ariste A, Schmidt W, Barthol P, Gandorfer A.
\newblock 2011, {Unnoticed Magnetic Field Oscillations in the Very Quiet Sun
  Revealed by SUNRISE/IMaX}.
\newblock \emph{ApJl}, 730:\penalty0 L37.
\newblock (\doi{10.1088/2041-8205/730/2/L37})

\bibitem[{Moreels} and {Van Doorsselaere}(2013)]{2013A&A...551A.137M}
{Moreels} MG, {Van Doorsselaere} T.
\newblock 2013, {Phase relations for seismology of photospheric flux tubes}.
\newblock \emph{A\&A}, 551:\penalty0 A137.
\newblock (\doi{10.1051/0004-6361/201219568})

\bibitem[{Moreels} et~al.(2013){Moreels}, {Goossens}, and {Van
  Doorsselaere}]{2013A&A...555A..75M}
{Moreels} MG, {Goossens} M, {Van Doorsselaere} T.
\newblock 2013, {Cross-sectional area and intensity variations of sausage
  modes}.
\newblock \emph{A\&A}, 555:\penalty0 A75.
\newblock (\doi{10.1051/0004-6361/201321545})

\bibitem[{Joshi} and {de la Cruz Rodrìguez}(2018)]{2018A&A...619A..63J}
{Joshi} J, {de la Cruz Rodrìguez} J.
\newblock 2018, {Magnetic field variations associated with umbral flashes and
  penumbral waves}.
\newblock \emph{A\&A}, 619:\penalty0 A63.
\newblock (\doi{10.1051/0004-6361/201832955})

\bibitem[Cavallini(2006)]{MScavallini06}
Cavallini F.
\newblock 2006, {IBIS: A New Post-Focus Instrument for Solar Imaging
  Spectroscopy}.
\newblock \emph{Sol. Phys.}, 236:\penalty0 415--439.
\newblock (\doi{10.1007/s11207-006-0103-8})

\bibitem[Reardon and Cavallini(2008)]{MSreardon08}
Reardon KP, Cavallini F.
\newblock 2008, {Characterization of Fabry-Perot interferometers and
  multi-etalon transmission profiles. The IBIS instrumental profile}.
\newblock \emph{A\&A}, 481:\penalty0 897--912.
\newblock (\doi{10.1051/0004-6361:20078473})

\bibitem[Rimmele(2004)]{2004SPIE.5490...34R}
Rimmele TR.
\newblock 2004,
\newblock , \emph{Advancements in Adaptive Optics}, 34--46volume 5490 of
  \emph{{\textbackslash}procspie}pages.
\newblock (\doi{10.1117/12.551764})

\bibitem[van Noort et~al.(2005)van Noort, Rouppe van~der Voort, and
  L{\"{o}}fdahl]{MSnoort05}
van Noort M, Rouppe van~der Voort L, L{\"{o}}fdahl MG.
\newblock 2005, {Solar Image Restoration By Use Of Multi-frame Blind
  De-convolution With Multiple Objects And Phase Diversity}.
\newblock \emph{Sol. Phys.}, 228:\penalty0 191--215.
\newblock (\doi{10.1007/s11207-005-5782-z})

\bibitem[Viticchi{\'{e}} et~al.(2010)Viticchi{\'{e}}, Del~Moro, Criscuoli, and
  Berrilli]{2010ApJ...723..787V}
Viticchi{\'{e}} B, Del~Moro D, Criscuoli S, Berrilli F.
\newblock 2010, {Imaging Spectropolarimetry with IBIS. II. On the Fine
  Structure of G-band Bright Features}.
\newblock \emph{ApJ}, 723:\penalty0 787--796.
\newblock (\doi{10.1088/0004-637X/723/1/787})

\bibitem[Stangalini et~al.(2018)Stangalini, Jafarzadeh, Ermolli, Erd{\'{e}}lyi,
  Jess, Keys, Giorgi, Murabito, Berrilli, and Moro]{Stangalini_2018}
Stangalini M, Jafarzadeh S, Ermolli I, Erd{\'{e}}lyi R, Jess DB, Keys PH,
  Giorgi F, Murabito M, Berrilli F, Moro DD.
\newblock 2018, Propagating spectropolarimetric disturbances in a large
  sunspot.
\newblock \emph{The Astrophysical Journal}, 869\penalty0 (2):\penalty0 110.
\newblock (\doi{10.3847/1538-4357/aaec7b}).
\newblock URL \url{https://doi.org/10.3847%2F1538-4357%2Faaec7b}

\bibitem[{Murabito} et~al.(2019){Murabito}, {Ermolli}, {Giorgi}, {Stangalini},
  {Guglielmino}, {Jafarzadeh}, {Socas-Navarro}, {Romano}, and
  {Zuccarello}]{2019ApJ...873..126M}
{Murabito} M, {Ermolli} I, {Giorgi} F, {Stangalini} M, {Guglielmino} SL,
  {Jafarzadeh} S, {Socas-Navarro} H, {Romano} P, {Zuccarello} F.
\newblock 2019, {Height Dependence of the Penumbral Fine-scale Structure in the
  Inner Solar Atmosphere}.
\newblock \emph{ApJ}, 873\penalty0 (2):\penalty0 126.
\newblock (\doi{10.3847/1538-4357/aaf727})

\bibitem[{Houston} et~al.(2020){Houston}, {Jess}, {Keppens}, {Stangalini},
  {Keys}, {Grant}, {Jafarzadeh}, {McFetridge}, {Murabito}, {Ermolli}, and
  {Giorgi}]{2020ApJ...892...49H}
{Houston} SJ, {Jess} DB, {Keppens} R, {Stangalini} M, {Keys} PH, {Grant} SDT,
  {Jafarzadeh} S, {McFetridge} LM, {Murabito} M, {Ermolli} I, {Giorgi} F.
\newblock 2020, {Magnetohydrodynamic Nonlinearities in Sunspot Atmospheres:
  Chromospheric Detections of Intermediate Shocks}.
\newblock \emph{ApJ}, 892\penalty0 (1):\penalty0 49.
\newblock (\doi{10.3847/1538-4357/ab7a90})

\bibitem[{Murabito} et~al.(2020){Murabito}, {Guglielmino}, {Ermolli},
  {Stangalini}, and {Giorgi}]{2020ApJ...890...96M}
{Murabito} M, {Guglielmino} SL, {Ermolli} I, {Stangalini} M, {Giorgi} F.
\newblock 2020, {Penumbral Brightening Events Observed in AR NOAA 12546}.
\newblock \emph{ApJ}, 890\penalty0 (2):\penalty0 96.
\newblock (\doi{10.3847/1538-4357/ab6664})

\bibitem[{MacBride} et~al.(2020){MacBride}, {Jess}, {Grant}, {Khomenko},
  {Keys}, and {Stangalini}]{2020arXiv200707904M}
{MacBride} CD, {Jess} DB, {Grant} SDT, {Khomenko} E, {Keys} PH, {Stangalini} M.
\newblock 2020, {Accurately constraining velocity information from spectral
  imaging observations using machine learning techniques}.
\newblock \emph{arXiv e-prints}, art. arXiv:2007.07904

\bibitem[{Stangalini, M.} et~al.(2012){Stangalini, M.}, {Giannattasio, F.},
  {Del Moro, D.}, and {Berrilli, F.}]{Stangalini2012}
{Stangalini, M.}, {Giannattasio, F.}, {Del Moro, D.}, {Berrilli, F.}
\newblock 2012, Three-minute wave enhancement in the solar photosphere.
\newblock \emph{A\&A}, 539:\penalty0 L4.
\newblock (\doi{10.1051/0004-6361/201118654}).
\newblock URL \url{https://doi.org/10.1051/0004-6361/201118654}

\bibitem[Torrence and Webster(1999)]{Torrence1999}
Torrence C, Webster PJ.
\newblock 1999, {Interdecadal Changes in the ENSO–Monsoon System}.
\newblock \emph{Journal of Climate}, 12\penalty0 (8):\penalty0 2679--2690.
\newblock ISSN 0894-8755.
\newblock (\doi{10.1175/1520-0442(1999)012<2679:ICITEM>2.0.CO;2}).
\newblock URL \url{http://adsabs.harvard.edu/abs/1999JCli...12.2679T}

\bibitem[Stangalini et~al.(2013)Stangalini, Solanki, Cameron, and
  Mart{\textbackslash}'{\textbackslash}inez~Pillet]{2013AA...554A.115S}
Stangalini M, Solanki SK, Cameron R,
  Mart{\textbackslash}'{\textbackslash}inez~Pillet V.
\newblock 2013, {First evidence of interaction between longitudinal and
  transverse waves in solar magnetic elements}.
\newblock \emph{A\&A}, 554:\penalty0 A115.
\newblock (\doi{10.1051/0004-6361/201220933})

\bibitem[Stangalini et~al.(2012)Stangalini, Giannattasio, Del~Moro, and
  Berrilli]{2012AA...539L...4S}
Stangalini M, Giannattasio F, Del~Moro D, Berrilli F.
\newblock 2012, {Three-minute wave enhancement in the solar photosphere}.
\newblock \emph{A\&A}, 539:\penalty0 L4.
\newblock (\doi{10.1051/0004-6361/201118654})

\bibitem[{Scherrer} et~al.(2012){Scherrer}, {Schou}, {Bush}, {Kosovichev},
  {Bogart}, {Hoeksema}, {Liu}, {Duvall}, {Zhao}, {Title}, {Schrijver},
  {Tarbell}, and {Tomczyk}]{2012SoPh..275..207S}
{Scherrer} PH, {Schou} J, {Bush} RI, {Kosovichev} AG, {Bogart} RS, {Hoeksema}
  JT, {Liu} Y, {Duvall} TL, {Zhao} J, {Title} AM, {Schrijver} CJ, {Tarbell} TD,
  {Tomczyk} S.
\newblock 2012, {The Helioseismic and Magnetic Imager (HMI) Investigation for
  the Solar Dynamics Observatory (SDO)}.
\newblock \emph{Sol. Phys.}, 275\penalty0 (1-2):\penalty0 207--227.
\newblock (\doi{10.1007/s11207-011-9834-2})

\bibitem[{Pesnell} et~al.(2012){Pesnell}, {Thompson}, and
  {Chamberlin}]{2012SoPh..275....3P}
{Pesnell} WD, {Thompson} BJ, {Chamberlin} PC.
\newblock 2012, {The Solar Dynamics Observatory (SDO)}.
\newblock \emph{Sol. Phys.}, 275\penalty0 (1-2):\penalty0 3--15.
\newblock (\doi{10.1007/s11207-011-9841-3})

\bibitem[{Lemen} et~al.(2012){Lemen}, {Title}, {Akin}, {Boerner}, {Chou},
  {Drake}, {Duncan}, {Edwards}, {Friedlaender}, {Heyman}, {Hurlburt}, {Katz},
  {Kushner}, {Levay}, {Lindgren}, {Mathur}, {McFeaters}, {Mitchell}, {Rehse},
  {Schrijver}, {Springer}, {Stern}, {Tarbell}, {Wuelser}, {Wolfson}, {Yanari},
  {Bookbinder}, {Cheimets}, {Caldwell}, {Deluca}, {Gates}, {Golub}, {Park},
  {Podgorski}, {Bush}, {Scherrer}, {Gummin}, {Smith}, {Auker}, {Jerram},
  {Pool}, {Soufli}, {Windt}, {Beardsley}, {Clapp}, {Lang}, and
  {Waltham}]{2012SoPh..275...17L}
{Lemen} JR, {Title} AM, {Akin} DJ, {Boerner} PF, {Chou} C, {Drake} JF, {Duncan}
  DW, {Edwards} CG, {Friedlaender} FM, {Heyman} GF, {Hurlburt} NE, {Katz} NL,
  {Kushner} GD, {Levay} M, {Lindgren} RW, {Mathur} DP, {McFeaters} EL,
  {Mitchell} S, {Rehse} RA, {Schrijver} CJ, {Springer} LA, {Stern} RA,
  {Tarbell} TD, {Wuelser} JP, {Wolfson} CJ, {Yanari} C, {Bookbinder} JA,
  {Cheimets} PN, {Caldwell} D, {Deluca} EE, {Gates} R, {Golub} L, {Park} S,
  {Podgorski} WA, {Bush} RI, {Scherrer} PH, {Gummin} MA, {Smith} P, {Auker} G,
  {Jerram} P, {Pool} P, {Soufli} R, {Windt} DL, {Beardsley} S, {Clapp} M,
  {Lang} J, {Waltham} N.
\newblock 2012, {The Atmospheric Imaging Assembly (AIA) on the Solar Dynamics
  Observatory (SDO)}.
\newblock \emph{Sol. Phys.}, 275\penalty0 (1-2):\penalty0 17--40.
\newblock (\doi{10.1007/s11207-011-9776-8})

\bibitem[{Cauzzi} et~al.(2008){Cauzzi}, {Reardon}, {Uitenbroek, H.},
  {Cavallini, F.}, {Falchi, A.}, {Falciani, R.}, {Janssen, K.}, {Rimmele, T.},
  {Vecchio, A.}, and {W\"oger, F.}]{Cauzzi2008}
{Cauzzi} G, {Reardon} K, {Uitenbroek, H.}, {Cavallini, F.}, {Falchi, A.},
  {Falciani, R.}, {Janssen, K.}, {Rimmele, T.}, {Vecchio, A.}, {W\"oger, F.}
\newblock 2008, The solar chromosphere at high resolution with ibis* - i. new
  insights from the ca854.2 nm line.
\newblock \emph{A\&A}, 480\penalty0 (2):\penalty0 515--526.
\newblock (\doi{10.1051/0004-6361:20078642}).
\newblock URL \url{https://doi.org/10.1051/0004-6361:20078642}

\bibitem[{Quintero Noda} et~al.(2017){Quintero Noda}, {Shimizu}, {Katsukawa},
  {de la Cruz Rodr{\'\i}guez}, {Carlsson}, {Anan}, {Oba}, {Ichimoto}, and
  {Suematsu}]{2017MNRAS.464.4534Q}
{Quintero Noda} C, {Shimizu} T, {Katsukawa} Y, {de la Cruz Rodr{\'\i}guez} J,
  {Carlsson} M, {Anan} T, {Oba} T, {Ichimoto} K, {Suematsu} Y.
\newblock 2017, {Chromospheric polarimetry through multiline observations of
  the 850-nm spectral region}.
\newblock (\doi{10.1093/mnras/stw2738})

\bibitem[{Baker} et~al.(){Baker}, {Stangalini}, {Valori}, and {et
  al.}]{Baker2020}
{Baker} D, {Stangalini} M, {Valori} G, {et al.}
\newblock Alfvénic perturbations in a sunspot chromosphere linked to highly
  fractionated plasma in the corona.
\newblock (in preparation)

\bibitem[Schunker and Cally(2006)]{2006MNRAS.372..551S}
Schunker H, Cally PS.
\newblock 2006, {Magnetic field inclination and atmospheric oscillations above
  solar active regions}.
\newblock \emph{MNRAS}, 372:\penalty0 551--564.
\newblock (\doi{10.1111/j.1365-2966.2006.10855.x})

\bibitem[{Khomenko} and {Collados}(2006)]{2006ApJ...653..739K}
{Khomenko} E, {Collados} M.
\newblock 2006, {Numerical Modeling of Magnetohydrodynamic Wave Propagation and
  Refraction in Sunspots}.
\newblock \emph{ApJ}, 653\penalty0 (1):\penalty0 739--755.
\newblock (\doi{10.1086/507760})

\bibitem[Cally and Goossens(2008)]{2008SoPh..251..251C}
Cally PS, Goossens M.
\newblock 2008, {Three-Dimensional MHD Wave Propagation and Conversion to
  Alfv{\{}{\'{e}}{\}}n Waves near the Solar Surface. I. Direct Numerical
  Solution}.
\newblock \emph{Sol. Phys.}, 251:\penalty0 251--265.
\newblock (\doi{10.1007/s11207-007-9086-3})

\bibitem[{Gary}(2001)]{2001SoPh..203...71G}
{Gary} GA.
\newblock 2001, {Plasma Beta above a Solar Active Region: Rethinking the
  Paradigm}.
\newblock \emph{Sol. Phys.}, 203\penalty0 (1):\penalty0 71--86.
\newblock (\doi{10.1023/A:1012722021820})

\bibitem[{Cally}(2011)]{2011ASInC...2..221C}
{Cally} PS.
\newblock 2011, {Alfv{\'e}n waves are easy: mode conversion in magnetic
  regions}.
\newblock In \emph{Astronomical Society of India Conference Series},
  221-227volume~2 of \emph{Astronomical Society of India Conference
  Series}pages

\end{thebibliography}

\end{document}